\documentstyle[preprint,pra,aps]{revtex}
\begin{document}
\draft
\title{\bf Interaction of instantons in a gauge theory
forcing their identical orientation}
\author{M.~Yu.~Kuchiev\cite{byline}}
\address{School of Physics, University of New South Wales,
Sydney, 2052, Australia.\\ email: kuchiev@newt.phys.unsw.edu.au }
\date{\today}
\maketitle
\begin{abstract}
A gauge theory model in which there exists a specific  interaction between
instantons is
considered. An effective action describing this interaction possesses a
minimum when
the instantons have  identical orientation.
The considered interaction might provide a phase transition into the state
where instantons have a preferred  orientation. This
phase of the gauge-field theory is important because it can give
the description of gravity in the framework of the gauge theory.
 \end{abstract}
\pacs{PACS: 04.60, 12.25}

The aim of this paper is to study a gauge theory model with an interaction
between instantons making their identical orientation preferable.
The interest to this problem is inspired by Ref.\cite{IAP} where
a possibility was found to describe effects of gravity in the framework of
a conventional $SO(4)$ gauge theory of Yang-Mills. In this approach
space-time is  supposed to be  basically flat.
On the basic level of the theory there are only  the usual for gauge theory
fields: gauge bosons as well as scalars and fermions interacting with the
gauge field. In order to exhibit effects of gravity the specific phenomenon,
the "condensation of polarized instantons and antiinstantons",
must take place in the vacuum of the $SO(4)$ gauge theory.
The instantons in one $su(2)$ subalgebra of $so(4)$ gauge algebra
must have the preferred
direction of orientation. The antiinstantons in the other
$su(2)\in so(4)$ subalgebra must also have their own proffered direction of
orientation. These orientations of condensates of
instantons and antiinstantons play
a role of the order parameter of the considered nontrivial phase of the
vacuum state.  The orientations of instantons
remain invariant under the
local gauge transformations, see Ref.\cite{Brown,ChW}. Therefore the possible
existence of the considered
nontrivial phase of the vacuum does not  contradict the
 gauge invariance,
which forbids any nontrivial state whose order parameter is noninvariant
under $local$ gauge transformations \cite{Eli}.

The most ``natural" way to look for the phase with polarised instantons
is to find an  interaction between the instantons
which forces any two instantons to have the identical orientation.
Then one can
expect  the system to undergo a  phase transition
into the state with polarized instantons.
Moreover, if one choose this scenario, then
the  antiinstantons of the
same $SU(2)$ group should remain noninteracting, thus resulting in the
absence of condensate of antiinstantons.
The problem is that in a pure gauge theory instantons do not interact.
There is the well-known interaction between instantons and antiinstantons
\cite{CDG} which depends on their orientation,
but there are exact multi-instantons solutions  with arbitrary
orientations of instantons \cite{AHDM}.
The main result of this paper is the $SU(2)$ gauge theory model
providing the  necessary interaction between a pair of
instantons  making their
identical orientation most probable and giving no interaction between
antiinstantons.

The qualitative description of the model is the following. Consider
$SU(2)$ gauge theory with scalars and fermions. Suppose that scalars
develop the  condensate. Suppose also that there is
an interaction between scalars and fermions which contains
scalar-pseudoscalar vertex $1-\gamma_5$. Then the right fermions are
influenced by the scalar condensate while the left fermions do not interact
with it. Consider now the gauge field created by several instantons.
The instantons are known to interact very strongly with the fermions. Therefore
 the fermions give a radiative correction to the effective action
which describes the instantons. Instanton influence
on the fermion field most strongly manifests itself in the fermionic
zero-modes \cite{tHo}.
In the pure gauge  theory with $n$ instantons there are $n$ degenerate
right-hand
zero-modes, as follows from the index theorem \cite{AtZ}. The existence
of the scalar condensate changes the situation drastically.
For this case the fermions are influenced by the vacuum scalar field
which obviously violates the condition of the index theorem
(the condition is that only the pure gauge field influences upon the fermions).
 Therefore the zero-modes can get splitting in the field of several instantons.
This splitting gives a contribution of  fermions into the
effective action describing the instanton field. The latter
 will be shown to depend on
the orientation of instantons. There is no such effect for antiinstantons
because their zero modes are left-handed and therefore they do not interact
with the scalar condensate, satisfy the condition of the index theorem
and thus exhibit no splitting.
The considered interaction arises as the
radiative one-fermion-loop correction
to the gauge field action.
Usually the radiative corrections  give only
the renormalization of  physical quantities. For the case considered
the correction  results in the new kind of interaction.

In order to formulate the model we are to chose the scalar condensate
in such a way that it creates the field applied to fermions
$V$  which satisfies
the condition $[\gamma \nabla , V]\ne 0$,
where $\nabla$ is the
covariant derivative $\gamma \nabla=
\gamma_\mu(\partial_\mu  - i A^a_\mu(x)T^a)$, and $T^a=\tau^a/2,~a=1,2,3$
are the generators of the gauge group. Otherwise no splitting
of the zero modes would arise. If one wishes to consider homogeneous,
$x$-independent field $V$, as is usual,
  then the only way to satisfy this condition is
to suppose that the field $V$ depends on the generators of the gauge group:
$V \sim \vec T\vec U(1-\gamma_5)$.  As a  result we are to
introduce into the problem some additional vector $\vec U$.
The  constant vector not only looks ugly but makes no good, as can be
verified.
We are to have the vector  whose averaged value is zero, but the averaged
values of its powers could play a role: $<U^a>=0,~<U^a U^b>=\vec U^2\delta_
{ab}, ~...$. The way to do it can only be provided by the additional
symmetry:
we are  to consider some ``additional'' $SU(2)$ group whose generators are
equal to the necessary vector $\vec U$. In this paper we will consider
this  $SU(2)$ group as a global one, it may be thought of as `` a flavour
group'' or ``a group of generations''.

Note that there is a clear and  interesting
analogy between the discussed construction and the
phenomenon of ferromagnetism.
The instantons might be compared with the atoms with nonzero spin,
the fermionic zero modes resemble the atomic outer electrons,
and  the scalar field which splits the zero modes plays a role very similar to
the crystal field which  creates  the conducting band. The problem
of instanton interaction, considered in the paper, looks
 similar to the problem
of the origin of exchange integral  in ferromagnetic theory.

Consider the $SU(2)$ gauge theory. Suppose that there are two generations
of fermionic fields with equal masses in the fundamental representation
of this gauge group.
We will treat them as a
doublet in the space of generations.
Suppose also that there are
three generations of scalars, - considered as a triplet in
the space of generations, -
in the vector representation of
$SU(2)$ gauge group.

Let us introduce an interaction between the scalars and the right-hand
fermions described by the Lagrangian
\begin{equation}\label{Lscalferm}
{\cal L} _{sf} (x)=  f  \psi_A^+(x)  \Phi_i(x) U^i_{AB} [(1-\gamma_5)/2]
\psi_B(x)~,
\end{equation}
where $f$ is the dimensionless constant of scalar-fermion interaction,
$\psi_A(x)$
is the fermion doublet, indexes $A,B = 1,2$ label the doublet
variables in the space
of generations, and $ \Phi_i(x), i=1,2,3$ is the triplet of scalar fields.
There is a freedom of choosing the matrixes $ U^i = U^i_{AB},~ i=1,2,3$
describing the coupling between different generations of fermions and scalars.
We choose these matrixes  be  the  triplet of generators of
rotations in the space of generations
\begin{equation} \label{vecU}
 U^i = U^i_{AB} = \sigma^i_{AB}/2,~ i=1,2,3~.
\end{equation}
The scalar fields $\Phi _i(x)$ are in the vector representation
 of the gauge $SU(2)$ group, therefore
$\Phi _i(x)=\Phi _{i,a}(x)T^a$.
Suppose now that  their nonlinear self-interaction  results in the
development of the
scalar condensate which has the  following form
\begin{equation}\label{scalcond}
 (\Phi _{i,a}(x))_{cond} = \phi \delta_{ia}~,
\end{equation}
where $\phi$ is a constant.  We see that the construction presented in
Eqs.(\ref{Lscalferm}),(\ref{vecU}),(\ref{scalcond})
results in the desired form  for the vacuum scalar field $V$
\begin{equation}\label{V}
V=  f \phi~ (\vec T  \vec U) (1-\gamma_5)/2 ~,
\end{equation}
which  influences on the right-hand fermions.
Note, that the Euclidean formulation is used.

Our goal is to calculate the
fermionic determinant $\det{(-i\gamma \nabla -i m -i V)}$,
when the gauge field $A^a_\mu(x)$ is created by $n=2$ instantons.
It is important that the determinant depends also on
the field $V$  (\ref{V})  created by
the scalar condensate.
Let us present it as
\begin{displaymath}
\det {(-i\gamma \nabla  -i m -i V)}=
\det {(-i\gamma  \nabla  -i m )}\det{(F)}~,
\end{displaymath}
where the first factor $\det {(-i\gamma \nabla -i m )}$ is
the determinant in a pure gauge field, and is not interesting
for our purposes, see discussion at the end of the paper.
Only the second  factor  is important
\begin{equation}\label{detGV}
\det{(F)}=\det{(1+GV)}=\exp{( -S_F)}~.
\end{equation}
Here $G$ is the propagator
of the fermions in the  gauge field,
$G=(\gamma \nabla  + m)^{-1}$,
and $S_F$ is the fermion correction to the gauge field action.
The instantons
create fermionic zero-modes playing  a
crucial role in the problem. Therefore, it is useful
to distinguish them in the fermion propagator.
With this purpose let us introduce
the projection operator onto the states of zero modes $P$.
It satisfies the conditions
$P^2=P~, ~(\gamma \nabla ) P = 0,~\mbox{Sp}(P)=2 n=4$.
Here $2n$ is the number of zero-modes.
Remember that we consider
$n=2$ instantons and 2 generations of fermions.
The propagator may be presented as $G~ = G_0 + G_1$ where
$G_0=P/m,~G_1=(1-P) (\gamma  \nabla +m)^{-1}(1-P)$.
To simplify   calculations let us consider the case of small
instantons,  so  that
the condition $m \rho \ll 1$,
where $\rho$ is an instanton radius, is fulfilled. Then the fermionic mass
$m$ may be considered as
a small parameter and we will put $m=0$ wherever it is possible.
In the limit $m=0$
the propagator of the   nonzero-modes was evaluated explicitly in
Ref.\cite{Brown}.
Using this propagator it is easy to verify that
$VG_1^{(m=0)}V=0$.
{}From this condition we find that nonzero modes are eliminated, only
zero modes contribute to (\ref{detGV})
\begin{equation}\label{zerodet}
\det{(F)}=\det{( 1+PV/m )}~.
\end{equation}
The operator $PV$ is presented by the
 finite dimensional, $2n\times 2n=4\times4$, matrix. Thus
the complicated fermionic functional
determinant is reduced for the case considered to the very simple one.

The wave functions of the zero modes are found  in terms
of the AHDM construction of Ref.\cite{AHDM}. With their help
the determinant in Eq.(\ref{detGV}) can be evaluated explicitly. The result
following from Eqs.(\ref{detGV}),(\ref{zerodet})
has a very simple form for
large separation between instantons
\begin{equation}\label{SFdilgaz}
S_F \approx \frac{f^2 \phi^2}{2 m^2}
\frac{\rho_1^2 \rho_2^2}{r^4}\sin^2{\gamma}~.
\end{equation}
Here
$\rho_1,\rho_2$ are the radii of the two instantons, $r$ is their
separation, $r > \rho_1,\rho_2$,  and
$\gamma$ is the angle beween the directions of the orientation of
the instantons  defined by the identity $\mbox{Re}(q_1 ^+q_2)=
\rho_1 \rho_2 \cos{\gamma}$, where $q_1,q_2$ are the quaternions
describing the orientations and radii of instantons \cite{AHDM}.

Note that if we consider the case when $\rho_1,\rho_2\ll f\phi$ then the scalar
condensate does not strongly disturb the instantons themselves.
At the same time, according to
Eq.(\ref{SFdilgaz}) there appears the  strong interaction between the
instantons making their identical orientation preferable.
There is no such interaction  between the antiinstantons.
One can reverse the situation changing the sign in front of $\gamma_5$
in Eq.(\ref{V}).
Then the antiinstantons interact and the instantons do not.
Evaluating this result we consider the correction
to the gauge field action given by the fermions in the one-loop approximation.
Note that we take into account
only those fermionic  loops
which are disturbed by the scalar condensate as well as by the gauge field.
We neglect  the loops of scalars and
gauge fields.
Nevertheless we can rely upon the obtained result.
Let us keep in mind that the instanton interaction
depends on the parameter $\zeta=f \phi/(2 m)$ which
has the fermion mass in the denominator. It
makes the found correction to the action to be
important for small $m$. Certainly  $\zeta$
should not be considered as a large parameter because the scalar condensate
might give a contribution to $m$ ($\delta m \sim f \phi $),
but one can consider both the scalar
condensate and the fermions mass  to be small, $ \phi,
{}~m \rightarrow 0$,
keeping their ratio constant: $\zeta =const$.
In this limit all
one-loop corrections to the action
omitted in the present calculations
are reduced to the ones
in a pure gauge theory
with the zero value of scalar condensate.
They are  recognised to give no interaction
between the instantons, their role in the problem of interest
is the renormalization
of the coupling constants and masses.
Therefore in the considered limit
the instanton-instanton interaction (\ref{SFdilgaz}) proves
to be valid.

The interaction (\ref{SFdilgaz})  between the instantons
allows one to address in the future
the problem of the phase transition into
the state in which the instantons are polarized and the antiinstantons
belonging to the same $SU(2)$ gauge group
are not (or inversely, the antiinstantons are polarised and
instantons are not).
To apply the results of this paper to the problem of gravity,
as it is formulated in \cite{IAP}, one has to resolve the
following problem. The construction considered in \cite{IAP}
required that the gauge field remained massless.
This is necessary, in particular, to
obtain massless gravitational waves in the theory.
This requirement is violated in the considered model:
there appears the mass of the gauge field $M_V=(3/4)g^2 \phi^2$.
One, heuristic, way to deal with this difficulty is
provided by the discussed above limit $\phi,
{}~m \rightarrow 0,~\zeta =const$, in which $M_V=0$.
Another possibility to avoid
the appearance of the  gauge field mass
is based upon a modification of the
considered model discussed elsewhere.

I thank V.V.Flambaum, C.J.Hamer, I.B.Khriplovich
and O.P.Sushkov for their  criticism of the obtained results
and G.F.Gribakin for critical comments on  the paper. A help
of L.S.Kuchieva in preparation of the manuscript is appreciated.
The financial support of Australian Research Council is acknowledged.


\begin{references}

\bibitem[^{(a)} ]{byline} On leave from: A.~F.~Ioffe
Physical-Technical Institute of Russian Academy of Sciences,
194021 St.Petersburg, Russian Federation.

\bibitem{IAP}
M.Yu.Kuchiev, $Europhys.Lett.$ {\bf 28} (1994) 539;

\bibitem{Brown}
L.~S.~Brown, R.~D.~Carlitz, D.~B.~Creamer, and C.~Lee,
$ Phys.~Rev.~D$ {\bf 17} (1978) 1583.

\bibitem{ChW}
N.~H.~Christ and E.~J.~Weinberg,
 $Phys.~Rev.~D$ {\bf 18} (1978) 2013.



\bibitem{Eli}
S.~Elitzur, $Phys.Rev.D$ {\bf 12} (1975) 3978.



\bibitem{CDG}
C.~Callan, K.~Dashen, and D.~Gross, $Phys.~Rev.~D$ {\bf 17} (1978) 2717;
{\it ibid}  {\bf 19} (1979) 1826.

\bibitem{AHDM}
M.~F.~Atiah, N.~J.~Hitchin, V.~G.~Drinfield, and Y.~I.~Manin,
$Phys.~Lett.A$ {\bf 65} (1978) 185.

\bibitem{tHo}
G.~'t~Hooft, $Phys.~Rev.~D$ {\bf 14}  (1976) 3432.

\bibitem{AtZ}
M.~F.~Atiah, I.M.Singer, Ann.Math. {\bf 87} (1968) 484.



\end{references}
\end{document}